# Bone Texture Analysis for Prediction of Incident Radiographic Hip Osteoarthritis Using Machine Learning: Data from the Cohort Hip and Cohort Knee (CHECK) study

Jukka Hirvasniemi[1,2], Willem Paul Gielis[2], Saeed Arbabi[3], Rintje Agricola[4], Willem Evert van Spil[5], Vahid Arbabi[2,6,7], Harrie Weinans[2,4,6]

[1]Center for Machine Vision and Signal Analysis, Faculty of Information Technology and Electrical Engineering, University of Oulu, Oulu, Finland
[2]Department of Orthopedics, University Medical Center Utrecht, Utrecht, The Netherlands
[3]Department of Computer Engineering, Faculty of Engineering, University of Zabol, Zabol, Iran
[4]Department of Orthopaedics, Erasmus University Medical Center, Rotterdam, the Netherlands
[5]Department of Rheumatology & Clinical Immunology, University Medical Center Utrecht, Utrecht, The Netherlands
[6]Department of Biomechanical Engineering, Delft University of Technology, Delft, The Netherlands
[7]Department of Mechanical Engineering, Faculty of Engineering, University of Birjand, Birjand, Iran

j.hirvasniemi@erasmusmc.nl

**Abstract.** Our aim was to assess the ability of radiography-based bone texture parameters in proximal femur and acetabulum to predict incident radiographic hip osteoarthritis (rHOA) over a 10 years period. Pelvic radiographs from CHECK (Cohort Hip and Cohort Knee) at baseline (987 hips) were analyzed for bone texture using fractal signature analysis in proximal femur and acetabulum. Elastic net (machine learning) was used to predict the incidence of rHOA (Kellgren-Lawrence grade (KL) $\geq$ 2 or total hip replacement (THR)), joint space narrowing score (JSN, range 0-3), and osteophyte score (OST, range 0-3) after 10 years. Performance of prediction models was assessed using the area under the receiver operating characteristic curve (ROC AUC). Of the 987 hips without rHOA at baseline, 435 (44%) had rHOA at 10-year follow-up. Of the 667 hips with JSN grade 0 at baseline, 471 (71%) had JSN grade $\geq$ 1 at 10-year follow-up. Of the 613 hips with OST grade 0 at baseline, 526 (86%) had OST grade $\geq$ 1 at 10-year follow-up. AUCs for the models including age, gender, and body mass index to predict incident rHOA, JSN, and OST were 0.59, 0.54, and 0.51, respectively. The inclusion of bone texture parameters in the models improved the prediction of incident rHOA (ROC AUC 0.66 and 0.71 when baseline KL was also included in the model) and JSN (ROC AUC 0.62), but not incident OST (ROC AUC 0.53). Bone texture analysis provides additional information for predicting incident rHOA or THR over 10 years.

# 1 Introduction

Plain radiography is a cheap, fast and widely available imaging method for osteoarthritis (OA). Bony changes can be clearly seen on plain radiographs and provide useful information about bone deformities, density, and structure. A plain radiograph is a projection (summation) through a three-dimensional structure and this is one main limitation of this imaging method. However, it has been shown that radiography-based bone texture is significantly related with the three-dimensional structure of bone [1-5].

Medical image analysis often involves interpretation of tissue appearance, e.g., smooth, grainy, rough, or homogenous. These image properties are related to the spatial arrangement of pixel intensities in images, i.e., image texture, and can be quantified using texture analysis [6]. Radiography-based texture analysis of the proximal femur has been applied for example in osteoporosis and in the assessment of femoral neck fracture risk [7, 8]. However, in OA research, the majority of studies analyzing bone texture are concentrated on the knee, using mostly fractal-based texture analysis methods [1, 9-13]. There is evidence that tibial trabecular bone texture can be used to predict both development and progression of OA as well as total knee replacement [14-21]. Only one study applied fractal signature analysis (FSA) on hip radiographs to quantify trabecular bone changes in subjects with prevalent hip OA and reported changes in fractal dimension of femoral head between baseline and 18 months follow-up [22]. However, the sample size of that study was relatively small (14 subjects) and the follow-up rather short.

Given the previous results showing that FSA can be applied on hip radiographs [22] and that bone density related parameters from dual energy X-ray absorptiometry (DXA) contribute to the risk and progression of hip OA [23-26], we hypothesize that radiography-based bone texture gives additional information in predicting the development of radiographic hip OA (rHOA). Consequently, our aim was to create a method for automated assessment of bone texture in proximal femur and acetabulum from plain hip radiographs and to assess the ability of these bone texture parameters to predict incident rHOA.

# 2 Subjects and Methods

## 2.1 Study cohort

Data from CHECK (Cohort Hip and Cohort Knee) cohort was used in this study [27]. CHECK is a prospective cohort study of 1002 subjects initiated to study the course of early knee and hip OA. Data was collected in ten medical centers in The Netherlands. Subjects were recruited by general practitioners and via advertisements. At baseline, subjects were aged 45-65 years, had first onset of pain or stiffness in hip(s) and/or knee(s), and had never or not longer than 6 months ago consulted a physician for these complaints. Subjects with a pathological condition other than early OA that could explain symptoms were excluded. The study was approved by medical ethics committees of all ten participating centers and written informed consent was obtained from all participants.

Plain pelvic anterior-posterior radiographs and clinical data at baseline and 10-year follow-up were used in the current study. Subjects with missing data (radiographs, demographics, clinical examination), KL grade ≥ 2 at baseline, and/or with insufficient radiograph quality (artefacts or underexposed) were excluded (Figure 1). As such, the final subset for assessing incident rHOA (KL grade ≥ 2) or total hip replacement (THR) within the period from baseline to 10 years included 987 hips (Table 1).

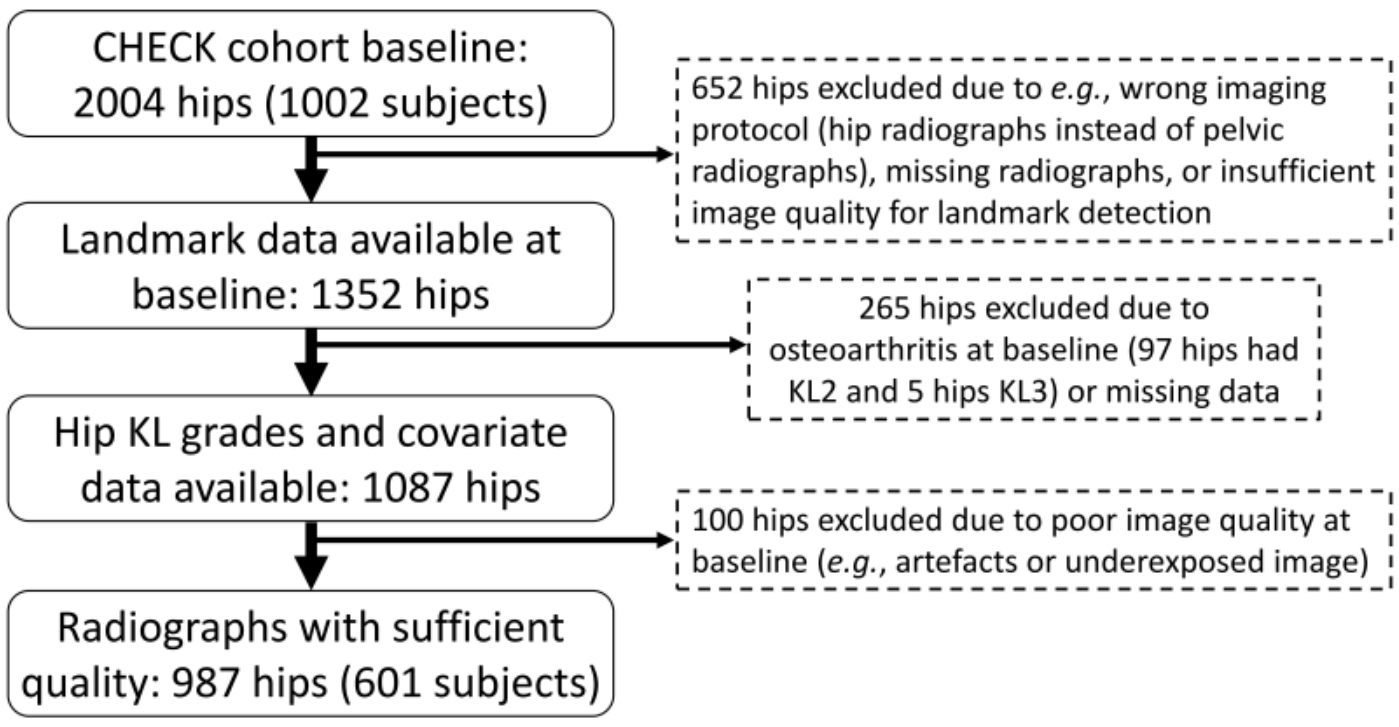

**Figure 1.** Flowchart of the selection of CHECK subjects and hips for the current study.

### 2.2 Acquisition and grading of the radiographs

Weight-bearing, anterior-posterior pelvic radiographs were acquired according to a standardized protocol. A wedge was used to assure 15-degree internal rotation in feet. The source – detector distance was 120 cm, and the X-ray beam was centered on the superior part of the pubic symphysis.

Hips were classified according to the KL grading scale [28] at baseline and 10-year follow-up. Superior and medial joint space narrowing (JSN) and superior and inferior osteophytes (OST) in acetabulum/femur were classified according to the Osteoarthritis Research Society International (OARSI) grading scale [29]. The scale for classifying the changes in JSN and OST was from 0 (normal) to 3 (severe change) [29]. Highest JSN and OST grades of the analyzed regions were used in the analyses. Table 1 summarizes the distribution of the KL, JSN, and OST grades of included hips at baseline and 10-year follow-up.

### 2.3 Selection of regions of interests

Prior to extraction of the regions of interests (ROIs), all images were resampled to have the same pixel size based on the smallest femoral head diameter (measured in pixels) on the data. Bicubic interpolation was used to ensure comparability of the structural parameters, without producing as much artefacts as bilinear or nearest neighbor interpolation algorithms. The resampling was also needed because part of the baseline radiographs were digitized and saved in TIFF images (501 hip images were in DICOM and 486 in TIFF format) and the actual pixel size on the detector was not available. After resampling, to assess bone texture from the radiographs, 41 circular ROIs with 70 pixels diameter were extracted from femoral head and acetabulum (25 on femoral head and 16 on acetabulum) (Figure 2). Although previous studies have typically used rectangular ROIs, circular ROIs were used in the current study to better cover femoral head and acetabular area and to enable bone texture assessment in many different directions inside the ROI without losing pixels when rotating the ROI. ROI selection procedure was based on fourteen out of seventy-five landmarks, which were manually placed on the proximal femur and pelvis in a previous study (Figure 2) [30]. Two circles were fitted in femoral head and acetabulum for ROI placement using Least Squares Optimization algorithm, which calculates the center and radius of best fitting circle in an iterative process. Locations of the ROIs were determined after a robustness assurance step which guarantees that the same ROI number selects the same corresponding pixels on images despite the size and rotation differences among them. Locations of the ROI1, ROI17, ROI25, and ROI26 were defined based on the center of the femoral head and the second-most lateral landmark on the acetabular rim (Figure 2). Other ROIs were automatically placed based on the locations of those ROIs.

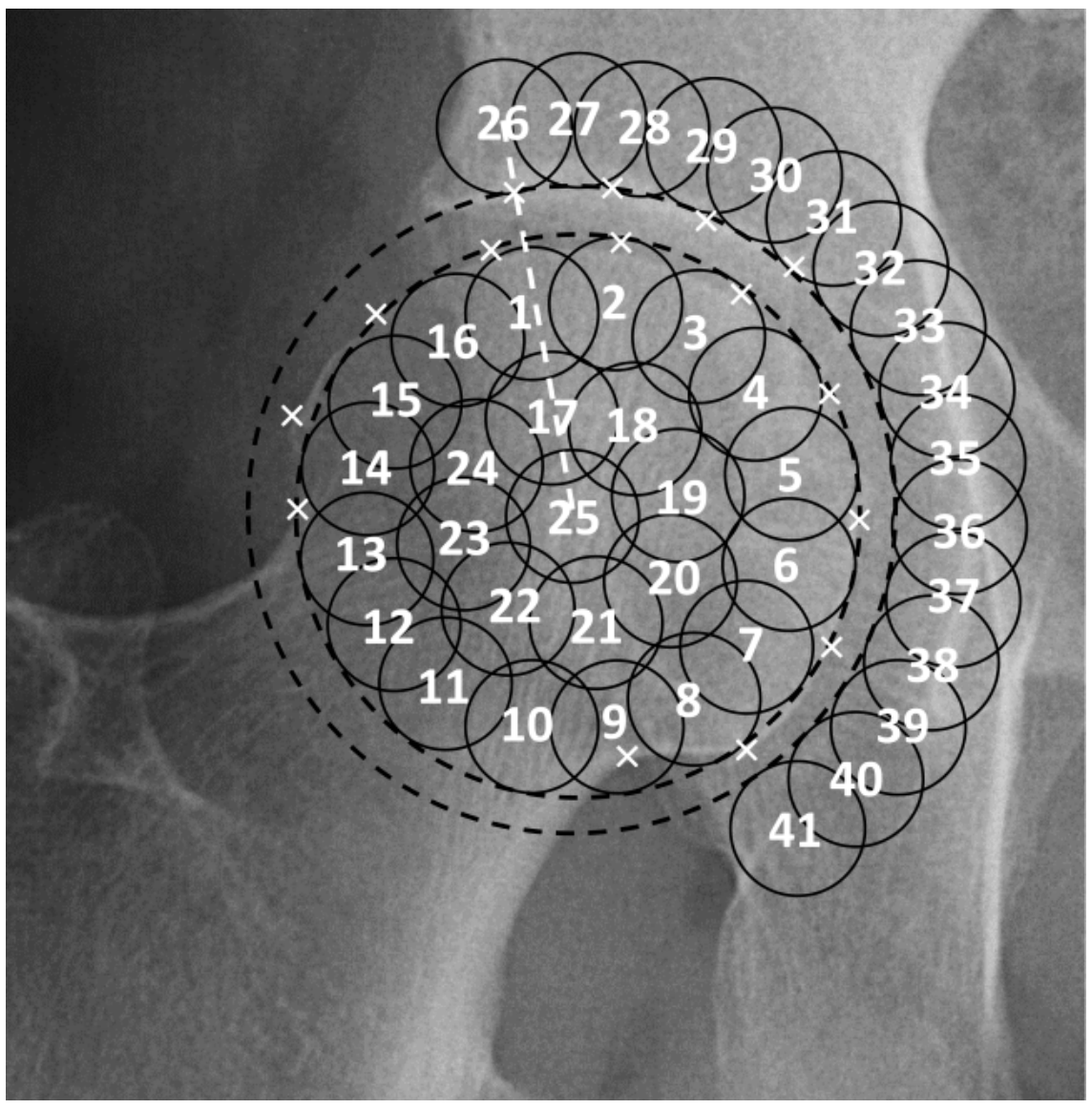

**Figure 2**. Location of regions of interest (ROIs). Landmarks that were used when fitting circles to the femoral head and acetabulum are indicated with white “x”. The white dashed line shows the centers of the femoral head, ROI1, ROI25 and ROI26. Other ROIs were automatically placed based on the locations of those ROIs.

## 2.4 Bone texture analysis

Before texture analysis, images were median filtered with a 3x3 pixels filter to remove high-frequency noise and grayscale values were expanded to full dynamic range (0 – 255). Bone texture was assessed using the FSA method [1, 10, 11]. FSA produces fractal dimension values that are related to the roughness and complexity of the image. To calculate the fractal dimensions, the image was dilated and eroded with a rod-shaped, horizontally oriented, one-pixel wide structuring element. After that, the volume, *V*, between dilated and eroded images was calculated. Calculations were repeated by varying the element length r from 2 to 5 pixels. The surface area, *A(r)*, was obtained from the Equation 1:

$$A(r) = (V(r)-V(r-1))/2 \tag{1}$$

Subsequently, a log-log plot was constructed by plotting log of *A(r)* against log of *r*. Finally, the fractal dimension was estimated by the slope of regression line that fitted the points in the log-log plot. High fractal dimension values are associated with high complexity of the image, whereas low complexity results in low fractal dimension values.

Because the orientation of the bone structures in femoral head and acetabulum varies, we assessed fractal dimensions in 18 different angles, i.e. from 0 to 170 degrees with 10 degrees increments. To reduce the number of fractal dimension values (18 values for 41 ROIs = 738 features per hip), minimum ($FD_{min}$) and maximum ($FD_{max}$) fractal dimension values and their respective angles ($Angles_{min}$, $Angles_{max}$) per ROI were selected. Consequently, 164 texture features were used in the analyses.

## 2.5 Statistical analyses

Logistic regression was used to assess the association of each baseline covariate (age, gender, body mass index (BMI), and baseline KL grade) as well as bone texture parameter with incident rHOA (KL grade $\geq$ 2), incident JSN (JSN grade $\geq$ 1), and incident OST (OST grade $\geq$ 1). Unadjusted odds ratios (OR) and 95% confidence interval of the univariate models were reported.

To prevent overfitting, machine learning was used for dimensionality reduction and to assess the predictive ability of the bone texture parameters and baseline covariates. For the dimensionality reduction and prediction, a regularized logistic regression method called elastic net was used [31, 32.] The elastic net linearly combines the L1 and L2 penalties of lasso and ridge regression methods [31, 32]. The samples were randomly divided into a training and validation set (790 hips, 80% of the data) and a hold-out test set (197 hips, 20% of the data) by stratifying the proportion of the controls and subjects with incident rHOA at follow-up in each set. To optimize the ratio of the L1 and L2 penalties ($\alpha$) and the strength of the penalty parameter ($\lambda$) of the elastic net, 10-fold cross-validation was performed. When $\alpha$ is close to zero, the elastic net approaches ridge regression, while when $\alpha$ is 1, lasso regression is performed. After optimizing elastic net parameters, the predictive ability of 1) covariate model, 2) texture feature model, 3) covariate + baseline KL model, and 4) model with covari-ates, baseline KL, and texture features combined in the test set were assessed using area under the receiver operating characteristics curve (ROC AUC). Analyses were repeated to predict incident rHOA among subjects with only KL0 or KL1 at baseline separately. Furthermore, incident JSN (JSN ≥ 1) among subjects with JSN grade 0 at baseline and incident OST (OST grade ≥ 1) among subjects with OST grade 0 at baseline were predicted with the elastic net. To remove the potential effect of imaging center to fractal dimension values, the parameters were standardized with mean and standard deviation values of the center where the imaging was performed ($z = (x - \mu)/SD$, where x is the value of each measurement, µ and SD are the average and standard deviation of the parameter at the center where the imaging was performed). Statistical analyses were performed using R (version 3.1.2) software with Caret [33] (version 6.0), pROC [34] (version 1.8) and glmnet [31] (version 2.0) packages.

## 3 Results

Of the 987 hips without rHOA at baseline, 435 (44%) had developed incident rHOA (KL ≥ 2 or THR) at10-year follow-up (Table 1). Of the 667 hips with JSN grade 0, 471 (71%) had JSN grade ≥ 1 or THR at 10-year follow-up. Of the 613 hips with OST grade 0, 526 (86%) had OST grade ≥ 1 or THR at 10-year follow-up.

The univariate logistic regression models for covariates showed that age (OR: 1.05), gender (OR: 0.61), and baseline KL grade (OR: 4.21) were significantly related to incident rHOA at the 10-year follow-up, but BMI was not (Table 2). When looking at the univariate unadjusted texture parameter models, ORs for minimum fractal dimension ($FD_{min}$) parameter models were statistically significant in 18/41 ROIs (OR range: 0.72 – 1.14), the maximum fractal dimension ($FD_{max}$) parameter in 13/41 ROIs (OR range: 0.80 – 1.15), the angle associated to $FD_{min}$ ($Angles_{min}$) in 8/41 ROIs (OR range: 0.95 – 1.03), and the angle associated to $FD_{max}$ ($Angles_{max}$) in 16/41 ROIs (OR range: 0.92 – 1.04) (Table 3). For incident JSN grade ≥ 1, ORs for age (OR: 1.04) and 42 texture parameters (OR range: 0.75 – 1.33) were significant (Tables 6 and 7). For incident OST grade ≥ 1, ORs for 15 texture parameters (OR range: 0.79 – 1.34) were significant (Tables 8 and 9).

**Table 2.** Odds ratios (95% confidence interval) of the univariate covariate models to assess incident rHOA (KL ≥ 2) or THR.

| Predictor | Odds ratio |
|---|---|
| Age (years) | 1.05 (1.03 – 1.08) |
| Female gender | 0.61 (0.45 – 0.83) |
| Baseline KL grade | 4.21 (3.26 – 5.46) |
| Body mass index (kg/m$^2$) | 1.00 (0.97 – 1.03) |

The selected elastic net parameters and ROC AUC values for the covariate model, texture model, covariate + baseline KL model, and for the combined covariate, baseline KL, and texture feature model from 10-fold cross-validation are shown in Table 4. The model that included covariates, baseline KL, and texture features had the high-est ROC AUC (0.72) in cross-validation. The variables that were selected based on the 10-fold cross-validation of the elastic net are listed in Table 5 and visualized in Figure 3. BMI was not selected in any of the models by the algorithm.

**Table 4.** Selected λ and α parameters for the elastic net from 10-fold cross-validation and areas under the receiver operating characteristics curve (ROC AUC) for predicting incidence of rHOA ($KL \geq 2$) or THR with the covariate, texture, covariate + baseline KL, and texture + covariate + baseline KL models.

| Model | Selected λ | Selected α | ROC AUC in validation |
|---|---|---|---|
| Covariates (age, gender, body mass index) | 0.082 | 0.45 | 0.60 |
| Texture features | 0.119 | 0.15 | 0.66 |
| Covariates + baseline KL | 0.247 | 0.25 | 0.69 |
| Texture features + covariates + baseline KL | 0.178 | 0.15 | 0.72 |

When assessing the performance of the optimized elastic net models in test set, the combined covariate, baseline KL, and bone texture feature model had the highest AUC (0.71 [95% CI: 0.63 – 0.78]). ROC curves for the 1) covariate, 2) texture, 3) combined covariates and baseline KL, and 4) combined covariate, baseline KL, and texture feature models in the test set are shown in Figure 4.

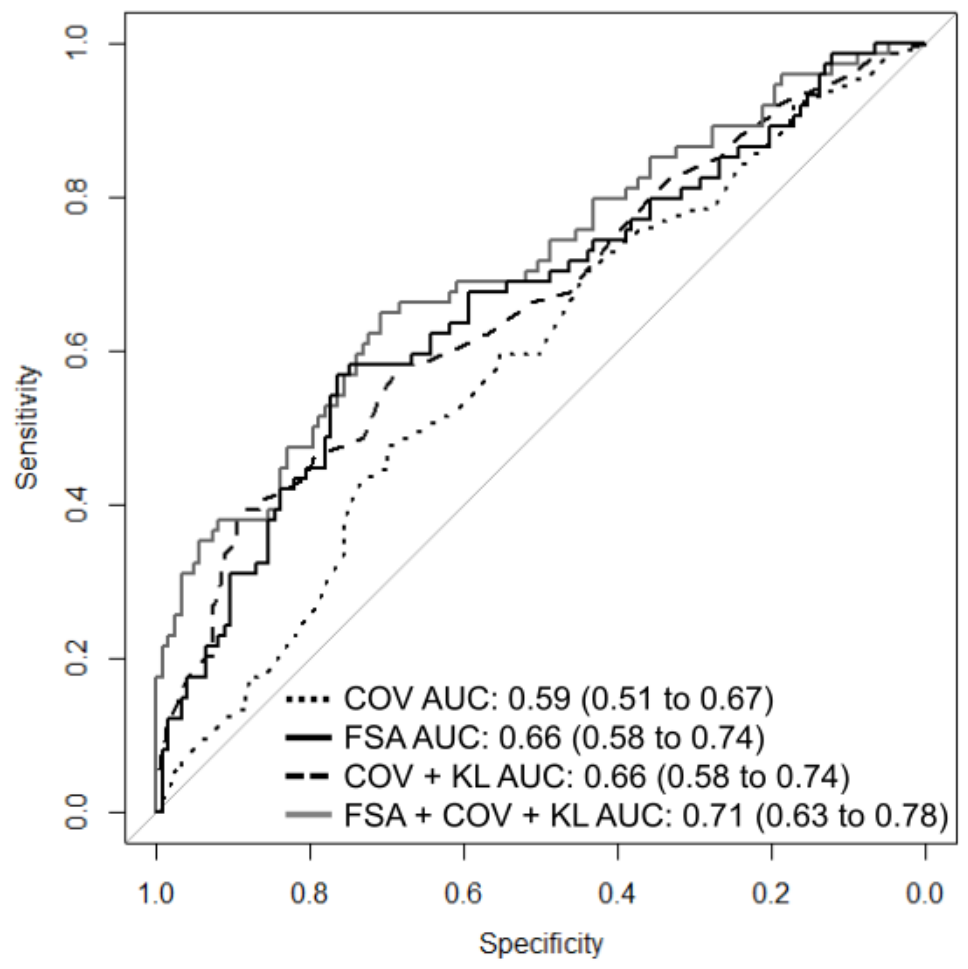

**Figure 4.** Receiver operating characteristics (ROC) curves and respective area under the curve (AUC) values for predicting incident rHOA (KL ≥ 2) or THR using 1) covariates (age, gender and body mass index), 2) texture parameters from fractal signature analysis (FSA), 3) covariates and baseline KL grade, and 4) texture parameters combined with covariates and KL grade.

When assessing subjects with KL0 or KL1 at baseline separately, adding the texture parameters improved model performance only among KL1 subjects in the test set (ROC AUC value increased from 0.54 to 0.67, Figures 5 and 6). For analyses in subjects with KL0 at baseline, ROC AUC values in cross-validation were 0.58, 0.60, and 0.61 for covariate, texture, and covariate + texture models, respectively. For analyses in subjects with KL1 at baseline, ROC AUC values in cross-validation were 0.67, 0.71, and 0.73 for covariate, texture, and covariate + texture models, respectively.

When predicting incident JSN in the test set, the combined texture and covariate model had the highest ROC AUC value of 0.62 (Figure 7). ROC AUC values in cross-validation were 0.56, 0.65, and 0.65 for covariate, texture, and covariate + texture models, respectively.

Selected models performed poorly when predicting incident OST in the test set (Figure 8). ROC AUC values in cross-validation were 0.56, 0.51, and 0.51 for covariate, texture, and covariate + texture models, respectively.

## 4 Discussion

In this study, we created a method for the assessment of bone texture in proximal femur and acetabulum from plain pelvic radiographs and assessed the ability of bone texture to predict incident rHOA or THR. Fractal dimension was measured from 41 ROIs that were placed on femoral head and acetabulum. Inclusion of bone texture parameters in the prediction model increased the ROC AUC value of the model during cross-validation (from 0.69 to 0.72) and in the hold-out test set (from 0.66 to 0.71) as compared to the model with baseline patient characteristics and baseline KL grade.

As there were no previous data for the optimal location of ROIs for bone texture analysis in the hip, we decided to cover the whole femoral ROI and also incorporate the acetabulum in our analyses. As shown in a previous study for the knee, areas distal from the subchondral bone might also include relevant texture information [14]. Interestingly, the relevant ROIs for the prediction of rHOA in our analyses were either at or next to the principal compressive trabeculae or close to the joint space (Figure 3). Depending on the ROI and variable, either higher or lower values were predictive for rHOA. For example, higher values in FDmin in ROIs 32 and 33 and lower values in ROIs 2, 6, 19, 28, 29, 32, 33, and 41 were predictive for rHOA. Lower fractal dimension values are likely associated to trabecular thickening or a reduction in trabecular number [22]. Changes in the angles associated to the fractal dimensions indicate the changes in the orientation of the trabeculae within the ROI. For example higher $Angles_{max}$ values indicate that the maximum fractal dimension value ($FD_{max}$) was detected from higher angles. One possible explanation for the differences between the directions of the predictive values in the ROIs may be the adaptation of bone according to the daily loading of the joint. Some areas may experience higher loads whereas the loads may be reduced in other areas. Furthermore, subchondral bone sclerosis affect the values in areas that are near the joint space.

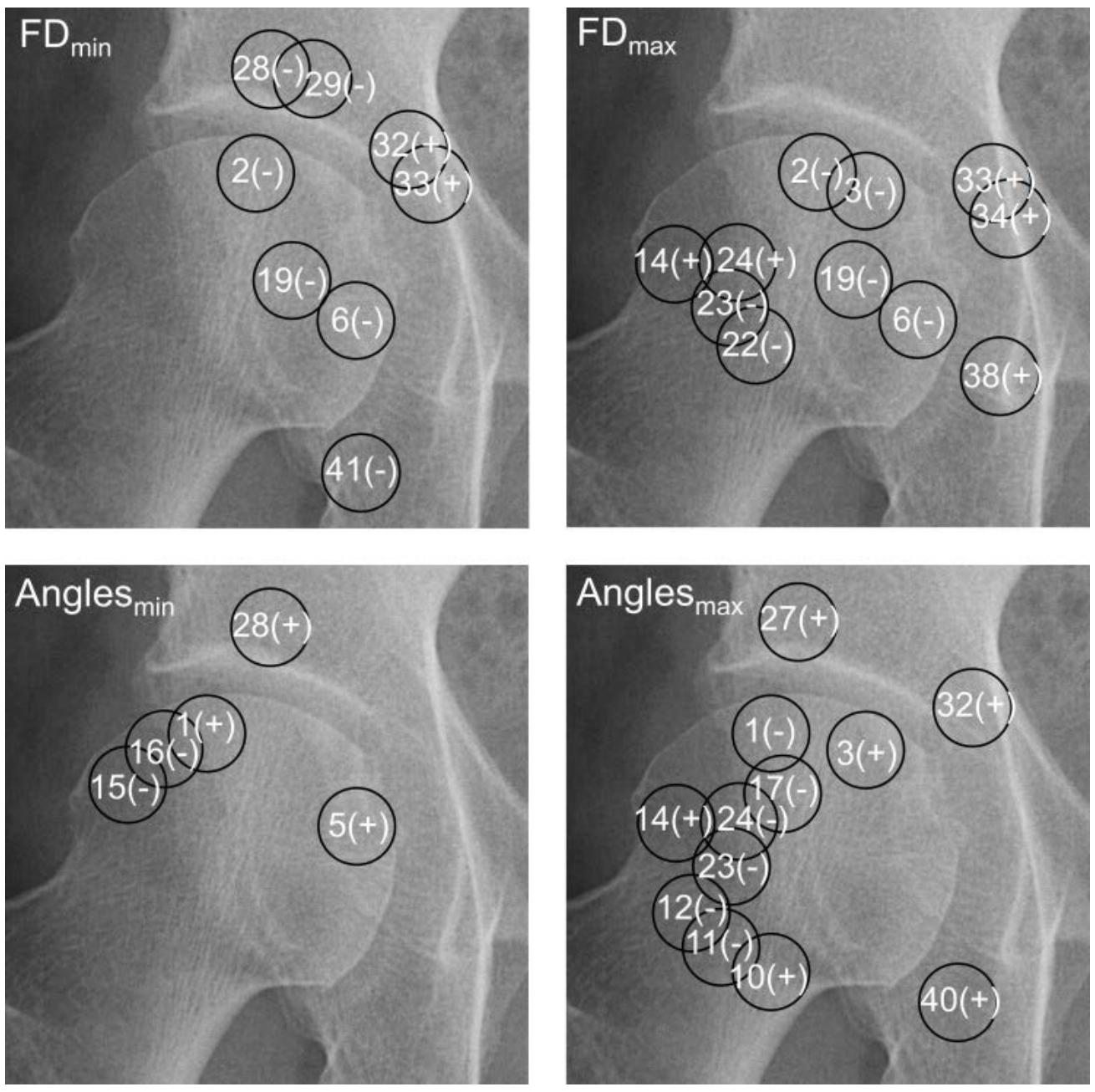

**Figure 3**. Location of regions of interest (ROIs) selected to the final elastic net model.

In a previous study assessing changes in FSA in OA hips, 14 subjects were followed 18 months. Fractal dimension of small and medium sized structures in the image were decreased during follow-up probably due to trabecular thickening or a reduction in trabecular number [22]. However, the relationship between FSA and development and/or progression of OA was not studied.

In contrast to the scarce assessment of associations between bone texture and hip OA, the association between tibial bone texture and knee OA has been described in many papers [14-16, 18-21]. ROC AUC values between 0.65 and 0.79 have been reported for models predicting progression of knee OA using bone texture and clinical covariates (e.g., age, BMI and gender) [14, 15, 18, 19]. JSN was used for defining progression in these studies. In a study assessing the predictive ability of texture parameters for incident knee OA defined as KL≥2 at follow-up, a ROC AUC of 0.69 was reported for the model including bone texture and clinical covariates (age, BMI and gender) [20]. Associations of covariates and bone texture with incident rHOA in our study are in line with these results. However, the ability to predict incident JSN or OST in the hip was worse than that reported for the knee [20]. The ROC AUC to predict rHOA increased from 0.66 to 0.71 after including bone texture to the model that included covariates and KL grading. Relatively low increase in ROC AUC may be because baseline KL grade alone is already a quite strong predictor of rHOA. Bone texture does not directly provide information about JSN, whereas JSN affects directly to KL grade, as this is included in its definition. However, trabecular bone structure is not evaluated in KL grading whereas bone texture analysis provides information about that.

When the ability to predict incident rHOA was compared between subjects with KL0 and KL1 at baseline, ROC AUC values were higher for KL1 subjects (ROC AUC values for the full model: 0.60 vs 0.67). These results suggest that in KL0 subjects there might not yet be changes that can be captured with bone texture analysis and predict incident rHOA.

This study contains some limitations that need to be discussed. First, the relevance of bone changes to the OA disease process might differ between HOA phenotypes. In the current study, different phenotypes were mixed as there is no consensus on how to define OA phenotypes yet. Second, radiographic scoring is subjective, semi-quantitative, and a plain radiograph is a projection of 3-dimensional structure. Therefore, some OA changes may have been missed. Third, rHOA and THR were combined as an outcome, while they might be different. We decided to combine these outcomes due to low number of THR subjects. Fourth, ten different medical centers participated in the data collection and different X-ray machines were used for the imaging, which may have affected texture analysis. However, FSA has been shown to be robust to the changes in imaging settings (e.g., exposure and pixel size) [10]. Furthermore, to remove possible effects from differences between centers, FSA parameters were standardized within each center. We think that including data from multiple X-ray machines increases the generalizability of our results. Fifth, training, validation, and test sets were derived from CHECK and the model was not tested in another cohort. However, to reduce overfitting, the hold-out test set was not used in cross-validation and the optimal elastic net parameters were searched using 10-fold cross-validation.

In conclusion, bone texture analysis in proximal femur and acetabulum provides additional information when trying to predict incident rHOA or THR. Our results suggest that bone texture parameters could be valuable when building prediction tools for OA.

## Acknowledgments

The research leading to these results has received funding from the Academy of Finland (project 308165). The funding sources had no role in the study design, data collection or analysis, interpretation of data, writing of the manuscript, or in the decision to submit the manuscript for publication.

The CHECK-cohort study is funded by the Dutch Arthritis Foundation. Involved are: Erasmus Medical Center Rotterdam; Kennemer Gasthuis Haarlem; Leiden University Medical Center; Maastricht University Medical Center; Martini Hospital Groningen /Allied Health Care Center for Rheumatology and Rehabilitation Groningen; Medical Spectrum Twente Enschede /Ziekenhuisgroep Twente Almelo; Reade Center for Rehabilitation and Rheumatology; St.Maartens-kliniek Nijmegen; University Medical Center Utrecht and Wilhelmina Hospital Assen.

**Table 1.** Characteristics of the subjects that were included in the current study.

| Variable | Baseline | 10-year follow-up | Incident OA (KL ≥ 2) | | Incident JSN (JSN grade ≥ 1) | | Incident OST (OST grade ≥ 1) | |
|---|---|---|---|---|---|---|---|---|
| | | | Controls (*n* = 552) | OA (*n* = 435) | Controls (*n* = 196) | JSN (*n* = 471) | Controls (*n* = 87) | OST (*n* = 526) |
| Anthropometric variables | | | | | | | | |
| Age [years] | 55.7 (5.2) | | 55.0 (5.2) | 56.5 (5.1) | 55.1 (5.0) | 56.0 (5.1) | 55.3 (5.6) | 55.2 (5.1) |
| Body mass index [kg/m$^2$] | 25.9 (3.8) | | 25.9 (3.9) | 25.9 (3.6) | 25.5 (3.2) | 25.9 (3.8) | 25.5 (3.5) | 26.0 (3.8) |
| Gender: | | | | | | | | |
| Male | 160 (16.2%) | | 71 (12.9%) | 89 (20.5%) | 23 (11.7%) | 67 (14.2%) | 9 (10.3%) | 74 (14.1%) |
| Female | 827 (83.8%) | | 481 (87.1%) | 346 (79.5%) | 173 (88.3%) | 404 (85.8%) | 78 (89.7%) | 452 (85.9%) |
| KL grade distribution: | | | | | | | | |
| KL0 | 720 (72.9%) | 131 (13.3%) | | | | | | |
| KL1 | 267 (27.1%) | 421 (42.7%) | | | | | | |
| KL2 | | 389 (39.4%) | | | | | | |
| KL3 | | 15 (1.5%) | | | | | | |
| KL4 | | - | | | | | | |
| Total hip replacement | | 31 (3.1%) | | | | | | |
| JSN grade distribution: | | | | | | | | |
| 0 | 667 (100%) | 196 (29.4%) | | | | | | |
| 1 | | 438 (65.7%) | | | | | | |
| 2 | | 21 (3.1%) | | | | | | |
| 3 | | 2 (0.3%) | | | | | | |
| Total hip replacement | | 10 (1.5%) | | | | | | |
| OST grade distribution: | | | | | | | | |
| 0 | 613 (100%) | 87 (14.2%) | | | | | | |
| 1 | | 348 (56.8%) | | | | | | |
| 2 | | 167 (27.2%) | | | | | | |
| 3 | | 5 (0.8%) | | | | | | |
| Total hip replacement | | 6 (1.0%) | | | | | | |

KL = Kellgren-Lawrence, JSN = joint space narrowing, OST = osteophyte. All values are given as mean (standard deviation) or *n* (%).

**Table 3**. Odds ratios (95% confidence interval) of unadjusted univariate texture parameter models to assess incidence of rHOA ($KL \geq 2$) or THR.

| ROI | $FD_{min}$ | $FD_{max}$ | $Angles_{min}$ | $Angles_{max}$ |
|---|---|---|---|---|
| 1 | 0.86 (0.77 – 0.95)* | 0.87 (0.78 – 0.97)* | 1.03 (1.01 – 1.06)* | 0.91 (0.88 – 0.94)* |
| 2 | 0.79 (0.71 – 0.88)* | 0.82 (0.74 – 0.92)* | 1.00 (0.97 – 1.03) | 0.97 (0.95 – 0.99)* |
| 3 | 0.92 (0.83 – 1.03) | 0.89 (0.80 – 0.99)* | 0.97 (0.94 – 1.00) | 1.01 (0.98 – 1.03) |
| 4 | 1.03 (0.93 – 1.15) | 0.98 (0.88 – 1.09) | 0.97 (0.94 – 1.00) | 0.99 (0.97 – 1.02) |
| 5 | 1.00 (0.90 – 1.11) | 0.99 (0.89 – 1.10) | 0.96 (0.94 – 0.99)* | 0.99 (0.96 – 1.02) |
| 6 | 0.84 (0.76 – 0.94)* | 0.87 (0.78 – 0.97)* | 0.99 (0.97 – 1.01) | 0.99 (0.96 – 1.02) |
| 7 | 0.95 (0.86 – 1.06) | 0.98 (0.88 – 1.09) | 0.99 (0.97 – 1.01) | 1.00 (0.97 – 1.03) |
| 8 | 1.03 (0.93 – 1.14) | 1.05 (0.95 – 1.17) | 1.00 (0.98 – 1.01) | 0.97 (0.94 – 0.99)* |
| 9 | 0.99 (0.89 – 1.10) | 1.01 (0.91 – 1.13) | 0.98 (0.96 – 1.01) | 0.98 (0.96 – 1.00)* |
| 10 | 0.82 (0.74 – 0.91)* | 0.88 (0.79 – 0.98)* | 1.02 (0.99 – 1.05) | 0.96 (0.94 – 0.98)* |
| 11 | 0.90 (0.81 – 0.99)* | 0.96 (0.86 – 1.06) | 1.01 (0.98 – 1.04) | 0.94 (0.92 – 0.96)* |
| 12 | 0.93 (0.83 – 1.03) | 1.03 (0.93 – 1.14) | 1.02 (0.99 – 1.04) | 0.93 (0.91 – 0.96)* |
| 13 | 0.92 (0.83 – 1.03) | 1.04 (0.94 – 1.16) | 1.02 (0.99 – 1.04) | 0.96 (0.93 – 0.99)* |
| 14 | 1.03 (0.93 – 1.15) | 1.15 (1.04 – 1.28)* | 1.03 (1.01 – 1.05)* | 0.94 (0.91 – 0.98)* |
| 15 | 0.97 (0.88 – 1.08) | 0.99 (0.89 – 1.10) | 0.98 (0.97 – 1.00)* | 1.02 (0.98 – 1.06) |
| 16 | 1.06 (0.95 – 1.17) | 0.92 (0.82 – 1.02) | 0.95 (0.93 – 0.97)* | 0.96 (0.92 – 1.01) |
| 17 | 1.03 (0.92 – 1.14) | 1.01 (0.91 – 1.12) | 1.00 (0.98 – 1.03) | 0.92 (0.90 – 0.94)* |
| 18 | 0.90 (0.81 – 0.99)* | 0.91 (0.82 – 1.02) | 0.96 (0.93 – 0.99)* | 0.98 (0.96 – 1.01) |
| 19 | 0.81 (0.73 – 0.90)* | 0.85 (0.77 – 0.95)* | 1.03 (0.99 – 1.06) | 0.98 (0.95 – 1.00) |
| 20 | 0.92 (0.82 – 1.02) | 0.99 (0.89 – 1.10) | 1.02 (0.99 – 1.05) | 0.98 (0.96 – 1.00)* |
| 21 | 1.00 (0.90 – 1.11) | 0.92 (0.83 – 1.02) | 0.99 (0.96 – 1.03) | 0.98 (0.96 – 1.01) |
| 22 | 0.89 (0.80 – 0.99)* | 0.86 (0.77 – 0.95)* | 0.97 (0.93 – 1.00) | 0.95 (0.93 – 0.98)* |
| 23 | 0.88 (0.79 – 0.98)* | 0.90 (0.81 – 1.00) | 1.03 (0.99 – 1.06) | 0.94 (0.92 – 0.97)* |
| 24 | 1.05 (0.94 – 1.16) | 1.09 (0.98 – 1.21) | 1.01 (0.99 – 1.04) | 0.93 (0.90 – 0.96)* |
| 25 | 0.98 (0.88 – 1.09) | 0.98 (0.88 – 1.09) | 0.98 (0.93 – 1.03) | 0.94 (0.90 – 0.97)* |
| 26 | 0.93 (0.84 – 1.04) | 0.99 (0.89 – 1.10) | 1.03 (0.99 – 1.05) | 1.02 (0.98 – 1.05) |
| 27 | 0.82 (0.74 – 0.92)* | 0.85 (0.76 – 0.94)* | 1.01 (0.98 – 1.04) | 0.98 (0.96 – 0.99)* |
| 28 | 0.81 (0.73 – 0.90)* | 0.86 (0.77 – 0.96)* | 1.02 (0.99 – 1.05) | 1.00 (0.98 – 1.01) |
| 29 | 0.79 (0.71 – 0.88)* | 0.95 (0.85 – 1.05) | 0.98 (0.95 – 1.01) | 1.01 (0.98 – 1.03) |
| 30 | 0.84 (0.75 – 0.93)* | 0.97 (0.88 – 1.08) | 1.00 (0.98 – 1.02) | 0.98 (0.96 – 1.01) |
| 31 | 1.00 (0.90 – 1.11) | 0.96 (0.87 – 1.07) | 0.98 (0.95 – 1.00)* | 0.98 (0.95 – 1.01) |
| 32 | 1.12 (1.01 – 1.25)* | 0.94 (0.84 – 1.04) | 0.98 (0.96 – 1.00) | 1.03 (0.99 – 1.07) |
| 33 | 1.14 (1.03 – 1.27)* | 1.09 (0.98 – 1.21) | 0.98 (0.96 – 1.00) | 0.96 (0.90 – 1.03) |
| 34 | 1.07 (0.96 – 1.19) | 1.10 (0.99 – 1.22) | 1.00 (0.98 – 1.01) | 1.02 (0.92 – 1.13) |
| 35 | 0.99 (0.89 – 1.10) | 1.01 (0.90 – 1.12) | 1.02 (0.99 – 1.03) | 0.97 (0.88 – 1.07) |
| 36 | 1.00 (0.90 – 1.12) | 0.92 (0.83 – 1.02) | 1.02 (0.99 – 1.03) | 0.94 (0.83 – 1.05) |
| 37 | 1.03 (0.93 – 1.15) | 1.02 (0.92 – 1.13) | 1.01 (0.99 – 1.02) | 0.90 (0.79 – 1.03) |
| 38 | 1.04 (0.93 – 1.15) | 1.14 (1.02 – 1.26)* | 0.99 (0.97 – 1.00) | 0.99 (0.93 – 1.05) |
| 39 | 1.13 (1.02 – 1.26)* | 0.84 (0.75 – 0.93) | 0.99 (0.97 – 1.01) | 1.00 (0.97 – 1.04) |
| 40 | 0.89 (0.80 – 0.99)* | 0.83 (0.75 – 0.93)* | 0.97 (0.96 – 0.99)* | 1.04 (1.01 – 1.06)* |
| 41 | 0.72 (0.64 – 0.80)* | 0.80 (0.71 – 0.89)* | 0.98 (0.96 – 1.00) | 1.01 (0.99 – 1.03) |

*$p < 0.05$

**Table 5.** Variables in the final elastic net model to predict incident rHOA ($KL \geq 2$) or THR.

| Variable | Coefficient |
|---|---|
| Intercept | -0.72 |
| $FD_{min}$ ROI2 | -0.02 |
| $FD_{min}$ ROI6 | -0.03 |
| $FD_{min}$ ROI19 | -0.06 |
| $FD_{min}$ ROI28 | -0.01 |
| $FD_{min}$ ROI29 | -0.02 |
| $FD_{min}$ ROI32 | 0.01 |
| $FD_{min}$ ROI33 | 0.00 |
| $FD_{min}$ ROI41 | -0.07 |
| $FD_{max}$ ROI2 | -0.04 |
| $FD_{max}$ ROI3 | -0.01 |
| $FD_{max}$ ROI6 | -0.01 |
| $FD_{max}$ ROI14 | 0.02 |
| $FD_{max}$ ROI19 | -0.01 |
| $FD_{max}$ ROI22 | -0.06 |
| $FD_{max}$ ROI23 | -0.01 |
| $FD_{max}$ ROI24 | 0.01 |
| $FD_{max}$ ROI33 | 0.01 |
| $FD_{max}$ ROI34 | 0.04 |
| $FD_{max}$ ROI38 | 0.04 |
| $Angles_{min}$ ROI1 | 0.00 |
| $Angles_{min}$ ROI5 | 0.00 |
| $Angles_{min}$ ROI15 | -0.01 |
| $Angles_{min}$ ROI16 | -0.01 |
| $Angles_{min}$ ROI28 | 0.00 |
| $Angles_{max}$ ROI1 | -0.01 |
| $Angles_{max}$ ROI3 | 0.00 |
| $Angles_{max}$ ROI10 | 0.00 |
| $Angles_{max}$ ROI11 | -0.01 |
| $Angles_{max}$ ROI12 | -0.01 |
| $Angles_{max}$ ROI14 | 0.00 |
| $Angles_{max}$ ROI17 | -0.02 |
| $Angles_{max}$ ROI23 | -0.01 |
| $Angles_{max}$ ROI24 | -0.01 |
| $Angles_{max}$ ROI27 | 0.00 |
| $Angles_{max}$ ROI32 | 0.00 |
| $Angles_{max}$ ROI40 | 0.01 |
| Age | 0.02 |
| Baseline KL | 0.63 |
| Gender | -0.11 |

**Table 6.** Odds ratios (95% confidence interval) of unadjusted univariate covariate models to predict incident joint space narrowing (JSN grade ≥ 1).

| Predictor | Odds ratio |
|---|---|
| Age (years) | 1.04 (1.01 – 1.07) |
| Female gender | 0.80 (0.52 – 1.21) |
| Body mass index (kg/m$^2$) | 1.03 (0.99 – 1.07) |

**Table 7**. Odds ratios (95% confidence interval) of unadjusted univariate texture parameter models to predict incident joint space narrowing (JSN grade ≥ 1).

| ROI | $FD_{min}$ | $FD_{max}$ | $Angles_{min}$ | $Angles_{max}$ |
|---|---|---|---|---|
| 1 | 0.95 (0.82 – 1.10) | 0.93 (0.81 – 1.07) | 1.01 (0.98 – 1.03) | 0.90 (0.86 – 0.94)* |
| 2 | 0.86 (0.71 – 0.99)* | 1.03 (0.90 – 1.18) | 0.98 (0.94 – 1.02) | 0.96 (0.93 – 0.98)* |
| 3 | 0.99 (0.86 – 1.15) | 1.09 (0.95 – 1.26) | 0.94 (0.90 – 0.99)* | 1.00 (0.97 – 1.03) |
| 4 | 1.09 (0.95 – 1.26) | 1.15 (0.99 – 1.33) | 0.97 (0.92 – 1.01) | 0.98 (0.95 – 1.02) |
| 5 | 0.95 (0.83 – 1.10) | 0.91 (0.79 – 1.05) | 0.96 (0.92 – 0.99)* | 0.99 (0.94 – 1.03) |
| 6 | 0.81 (0.70 – 0.94)* | 0.86 (0.74 – 0.98)* | 0.99 (0.96 – 1.01) | 1.00 (0.96 – 1.04) |
| 7 | 0.95 (0.83 – 1.09) | 0.92 (0.80 – 1.06) | 1.01 (0.99 – 1.03) | 1.00 (0.97 – 1.04) |
| 8 | 1.29 (1.12 – 1.50)* | 1.24 (1.07 – 1.43)* | 0.96 (0.94 – 0.99)* | 1.02 (0.99 – 1.06) |
| 9 | 0.93 (0.81 – 1.07) | 1.03 (0.90 – 1.19) | 1.02 (0.98 – 1.05) | 0.98 (0.96 – 1.01) |
| 10 | 0.83 (0.72 – 0.95)* | 0.93 (0.81 – 1.07) | 1.05 (1.01 – 1.09)* | 0.97 (0.95 – 0.99)* |
| 11 | 1.01 (0.88 – 1.16) | 0.96 (0.83 – 1.10) | 1.03 (0.99 – 1.07) | 0.92 (0.89 – 0.95)* |
| 12 | 0.93 (0.81 – 1.08) | 0.96 (0.83 – 1.10) | 1.04 (1.01 – 1.06)* | 0.91 (0.87 – 0.95)* |
| 13 | 0.99 (0.86 – 1.14) | 1.11 (0.97 – 1.27) | 1.03 (1.00 – 1.05)* | 0.91 (0.87 – 0.95)* |
| 14 | 1.11 (0.96 – 1.29) | 1.16 (1.01 – 1.34)* | 1.03 (1.01 – 1.05)* | 0.98 (0.94 – 1.02) |
| 15 | 1.07 (0.94 – 1.23) | 1.07 (0.94 – 1.23) | 0.98 (0.96 – 1.00) | 1.03 (0.98 – 1.08) |
| 16 | 1.04 (0.91 – 1.20) | 0.98 (0.85 – 1.12) | 0.95 (0.92 – 0.98)* | 0.92 (0.86 – 0.98)* |
| 17 | 1.20 (1.04 – 1.39)* | 1.13 (0.98 – 1.30) | 1.02 (0.99 – 1.05) | 0.90 (0.86 – 0.93)* |
| 18 | 0.90 (0.78 – 1.04) | 1.00 (0.87 – 1.15) | 0.96 (0.91 – 1.01) | 0.97 (0.94 – 1.00) |
| 19 | 0.91 (0.78 – 1.05) | 0.84 (0.73 – 0.97)* | 1.07 (1.03 – 1.12)* | 0.98 (0.95 – 1.02) |
| 20 | 0.99 (0.86 – 1.14) | 0.96 (0.84 – 1.10) | 0.98 (0.94 – 1.02) | 1.02 (0.99 – 1.05) |
| 21 | 0.94 (0.82 – 1.07) | 0.90 (0.78 – 1.03) | 1.00 (0.96 – 1.05) | 0.96 (0.93 – 0.99)* |
| 22 | 0.88 (0.77 – 1.01) | 1.02 (0.89 – 1.17) | 1.03 (0.99 – 1.07) | 0.92 (0.90 – 0.95)* |
| 23 | 0.95 (0.83 – 1.09) | 1.03 (0.90 – 1.18) | 1.03 (0.99 – 1.06) | 0.91 (0.88 – 0.95)* |
| 24 | 1.03 (0.89 – 1.18) | 1.10 (0.96 – 1.26) | 1.03 (1.00 – 1.06)* | 0.92 (0.88 – 0.96)* |
| 25 | 1.03 (0.90 – 1.19) | 1.01 (0.88 – 1.16) | 1.04 (0.96 – 1.11) | 0.88 (0.84 – 0.92)* |
| 26 | 1.10 (0.95 – 1.27) | 1.17 (1.01 – 1.35)* | 1.05 (1.01 – 1.09)* | 1.01 (0.97–1.06 |
| 27 | 1.01 (0.88 – 1.16) | 0.95 (0.83 – 1.09) | 1.00 (0.96 – 1.05) | 0.98 (0.96 – 1.00) |
| 28 | 0.98 (0.85 – 1.13) | 0.85 (0.74 – 0.97)* | 0.98 (0.94 – 1.01) | 0.98 (0.97 – 1.00) |
| 29 | 0.87 (0.76 – 1.01) | 0.91 (0.80 – 1.05) | 0.99 (0.96 – 1.03) | 1.01 (0.98 – 1.05) |
| 30 | 0.96 (0.83 – 1.11) | 0.92 (0.80 – 1.05) | 0.99 (0.96 – 1.03) | 0.99 (0.96 – 1.02) |
| 31 | 0.96 (0.84 – 1.11) | 0.91 (0.79 – 1.04) | 0.98 (0.95 – 1.01) | 1.00 (0.96 – 1.03) |
| 32 | 1.05 (0.91 – 1.21) | 0.98 (0.85 – 1.13) | 0.97 (0.94 – 0.99)* | 0.97 (0.91 – 1.02) |
| 33 | 1.07 (0.92 – 1.24) | 1.08 (0.94 – 1.25) | 0.98 (0.95 – 1.00) | 0.92 (0.83 – 1.01) |
| 34 | 1.10 (0.96 – 1.28) | 1.00 (0.87 – 1.15) | 1.00 (0.98 – 1.02) | 0.93 (0.81 – 1.06) |
| 35 | 1.04 (0.90 – 1.20) | 0.88 (0.76 – 1.01) | 1.00 (0.98 – 1.02) | 0.99 (0.87 – 1.11) |
| 36 | 1.01 (0.88 – 1.17) | 0.89 (0.77 – 1.03) | 1.01 (0.99 – 1.03) | 1.02 (0.86 – 1.22) |
| 37 | 0.96 (0.84 – 1.11) | 1.05 (0.91 – 1.20) | 0.98 (0.96 – 1.00) | 0.87 (0.72 – 1.04) |
| 38 | 1.09 (0.94 – 1.26) | 1.33 (1.16 – 1.54)* | 0.98 (0.96 – 0.99)* | 1.05 (0.97 – 1.13) |
| 39 | 1.04 (0.90 – 1.20) | 1.04 (0.91 – 1.20) | 0.97 (0.95 – 1.00) | 1.05 (1.01 – 1.10)* |
| 40 | 0.78 (0.68 – 0.90)* | 0.75 (0.64 – 0.86)* | 1.02 (0.99 – 1.04) | 1.02 (0.99 – 1.05) |
| 41 | 0.79 (0.68 – 0.92)* | 0.88 (0.76 – 1.02) | 1.01 (0.98 – 1.04) | 1.01 (0.98 – 1.04) |

*$p < 0.05$

**Table 8.** Odds ratios (95% confidence interval) of the unadjusted univariate covariate models to assess incident osteophytes (OST grade ≥ 1).

| Predictor | Odds ratio |
|---|---|
| Age (years) | 1.00 (0.96 – 1.03) |
| Female gender | 0.70 (0.36 – 1.26) |
| Body mass index (kg/m$^2$) | 1.04 (0.99 – 1.09) |

**Table 9**. Odds ratios (95% confidence interval) of unadjusted univariate texture parameter models to assess incident osteophytes (OST grade ≥ 1).

| ROI | $FD_{min}$ | $FD_{max}$ | $Angles_{min}$ | $Angles_{max}$ |
|---|---|---|---|---|
| 1 | 0.99 (0.81 – 1.19) | 0.88 (0.72 – 1.07) | 0.97 (0.93 – 1.00) | 1.03 (0.97 – 1.09) |
| 2 | 1.08 (0.89 – 1.30) | 0.79 (0.65 – 0.95)* | 1.01 (0.96 – 1.06) | 1.02 (0.98 – 1.06) |
| 3 | 1.05 (0.86 – 1.26) | 0.88 (0.72 – 1.07) | 0.97 (0.91 – 1.03) | 1.02 (0.97 – 1.07) |
| 4 | 1.08 (0.89 – 1.31) | 0.99 (0.81 – 1.20) | 1.00 (0.95 – 1.06) | 1.03 (0.98 – 1.08) |
| 5 | 0.96 (0.80 – 1.16) | 0.97 (0.80 – 1.16) | 0.98 (0.94 – 1.03) | 1.02 (0.96 – 1.08) |
| 6 | 0.86 (0.70 – 1.05) | 0.81 (0.67 – 0.98)* | 1.02 (0.98 – 1.05) | 1.03 (0.97 – 1.09) |
| 7 | 0.98 (0.80 – 1.19) | 0.87 (0.72 – 1.04) | 0.97 (0.94 – 1.00) | 1.03 (0.99 – 1.08) |
| 8 | 0.89 (0.73 – 1.07) | 0.87 (0.71 – 1.05) | 1.01 (0.98 – 1.04) | 0.97 (0.93 – 1.02) |
| 9 | 1.01 (0.84 – 1.22) | 1.04 (0.86 – 1.26) | 0.99 (0.94 – 1.03) | 0.99 (0.96 – 1.02) |
| 10 | 0.83 (0.68 – 0.99)* | 0.94 (0.78 – 1.14) | 1.00 (0.94 – 1.05) | 0.97 (0.94 – 1.00) |
| 11 | 0.90 (0.73 – 1.11) | 0.88 (0.72 – 1.08) | 1.04 (0.99 – 1.09) | 1.00 (0.96 – 1.05) |
| 12 | 0.91 (0.75 – 1.10) | 0.96 (0.79 – 1.16) | 0.99 (0.94 – 1.03) | 1.00 (0.96 – 1.05) |
| 13 | 0.90 (0.74 – 1.09) | 1.13 (0.94 – 1.34) | 1.00 (0.97 – 1.04) | 1.00 (0.94 – 1.05) |
| 14 | 0.97 (0.79 – 1.18) | 1.00 (0.82 – 1.22) | 1.02 (0.99 – 1.05) | 1.05 (0.99 – 1.11) |
| 15 | 0.96 (0.79 – 1.15) | 1.01 (0.82 – 1.22) | 0.99 (0.96 – 1.02) | 1.07 (1.01 – 1.15)* |
| 16 | 0.99 (0.82 – 1.20) | 0.99 (0.82 – 1.20) | 0.99 (0.94 – 1.03) | 1.05 (0.97 – 1.13) |
| 17 | 0.84 (0.69 – 1.02) | 0.85 (0.70 – 1.03) | 0.99 (0.95 – 1.03) | 1.00 (0.95 – 1.05) |
| 18 | 0.94 (0.78 – 1.14) | 0.86 (0.71 – 1.04) | 0.99 (0.93 – 1.06) | 1.02 (0.98 – 1.07) |
| 19 | 0.85 (0.70 – 1.03) | 0.86 (0.71 – 1.04) | 1.09 (1.03 – 1.14)* | 1.02 (0.97 – 1.07) |
| 20 | 0.81 (0.66 – 0.99)* | 0.85 (0.70 – 1.02) | 1.00 (0.95 – 1.05) | 1.03 (0.99 – 1.08) |
| 21 | 0.82 (0.67 – 0.99)* | 0.93 (0.76 – 1.13) | 0.93 (0.86 – 0.99)* | 1.01 (0.97 – 1.07) |
| 22 | 0.84 (0.69 – 1.02) | 0.86 (0.70 – 1.04) | 0.97 (0.91 – 1.04) | 0.98 (0.94 – 1.03) |
| 23 | 0.81 (0.67 – 0.99)* | 0.91 (0.74 – 1.10) | 1.02 (0.98 – 1.06) | 1.03 (0.98 – 1.08) |
| 24 | 1.02 (0.84 – 1.23) | 1.09 (0.90 – 1.30) | 0.99 (0.95 – 1.04) | 1.02 (0.97 – 1.08) |
| 25 | 1.03 (0.85 – 1.25) | 1.08 (0.89 – 1.30) | 0.90 (0.79 – 0.99)* | 0.98 (0.92 – 1.03) |
| 26 | 1.01 (0.84 – 1.22) | 0.95 (0.79 – 1.15) | 1.05 (0.99 – 1.12) | 0.98 (0.92 – 1.04) |
| 27 | 0.92 (0.77 – 1.12) | 0.98 (0.81 – 1.18) | 1.05 (0.99 – 1.12) | 0.98 (0.95 – 1.01) |
| 28 | 0.89 (0.74 – 1.08) | 0.88 (0.72 – 1.06) | 1.01 (0.96 – 1.06) | 1.01 (0.98 – 1.04) |
| 29 | 0.92 (0.76 – 1.11) | 0.93 (0.77 – 1.12) | 0.97 (0.93 – 1.02) | 1.00 (0.96 – 1.05) |
| 30 | 0.95 (0.78 – 1.15) | 1.05 (0.86 – 1.27) | 0.99 (0.95 – 1.04) | 1.00 (0.96 – 1.05) |
| 31 | 0.89 (0.74 – 1.08) | 0.86 (0.71 – 1.04) | 0.99 (0.95 – 1.03) | 0.99 (0.95 – 1.04) |
| 32 | 1.04 (0.86 – 1.25) | 0.88 (0.73 – 1.07) | 0.99 (0.95 – 1.03) | 1.00 (0.92 – 1.07) |
| 33 | 0.99 (0.82 – 1.20) | 1.14 (0.94 – 1.38) | 1.03 (0.99 – 1.06) | 0.98 (0.85 – 1.11) |
| 34 | 1.08 (0.89 – 1.32) | 1.13 (0.94 – 1.37) | 1.00 (0.97 – 1.03) | 1.09 (0.92 – 1.28) |
| 35 | 0.98 (0.81 – 1.20) | 1.10 (0.91 – 1.32) | 1.00 (0.97 – 1.03) | 0.96 (0.81 – 1.13) |
| 36 | 1.02 (0.84 – 1.24) | 0.93 (0.77 – 1.12) | 1.05 (1.02 – 1.09)* | 0.93 (0.75 – 1.15) |
| 37 | 1.07 (0.88 – 1.30) | 0.79 (0.66 – 0.96)* | 1.03 (1.00 – 1.06)* | 0.94 (0.70 – 1.21) |
| 38 | 1.08 (0.88 – 1.32) | 1.09 (0.90 – 1.32) | 1.02 (0.99 – 1.05) | 0.87 (0.74 – 1.01) |
| 39 | 1.34 (1.08 – 1.69)* | 1.11 (0.92 – 1.35) | 0.99 (0.96 – 1.03) | 0.91 (0.84 – 0.98)* |
| 40 | 1.17 (0.97 – 1.42) | 1.04 (0.86 – 1.26) | 0.98 (0.94 – 1.02) | 1.00 (0.95 – 1.05) |
| 41 | 0.90 (0.73 – 1.09) | 0.91 (0.75 – 1.11) | 1.01 (0.97 – 1.04) | 1.00 (0.96 – 1.03) |

*$p < 0.05$

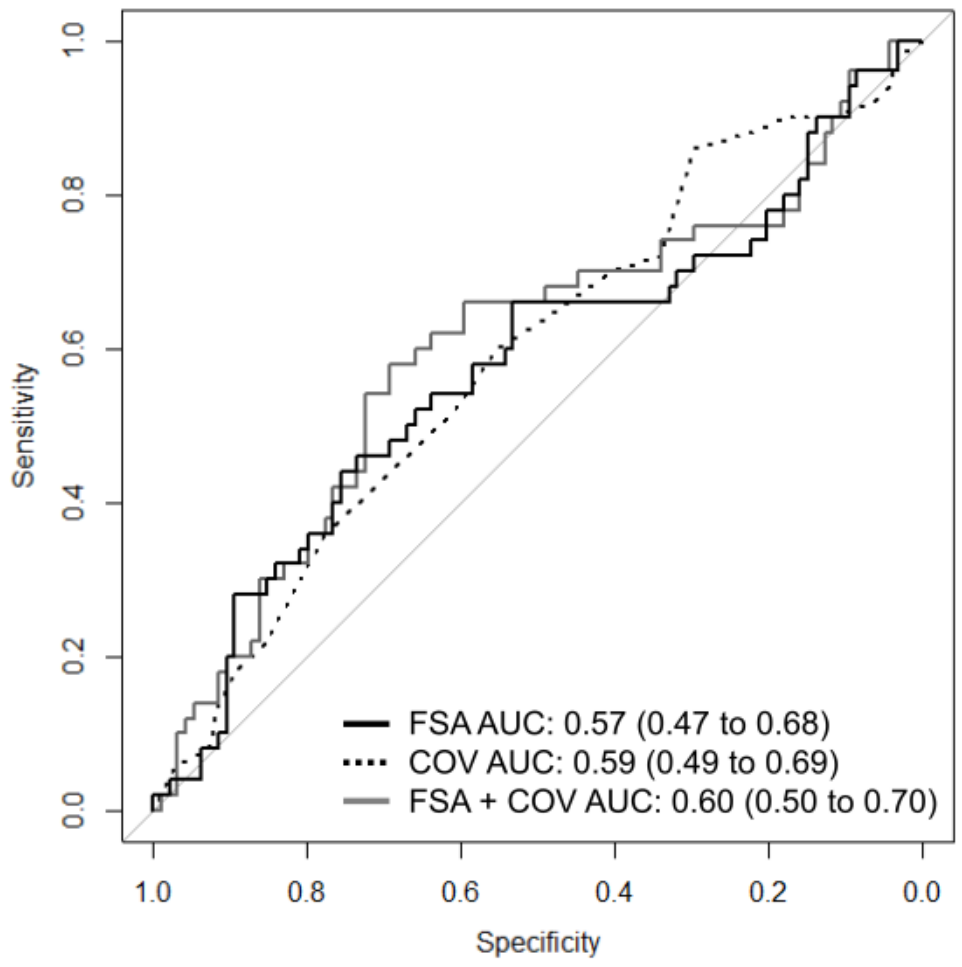

**Figure 5.** Receiver operating characteristics (ROC) curves and their respective area under the curve (AUC) values for predicting incident rHOA (KL ≥ 2) or THR among subjects with KL0 grade at baseline using 1) texture parameters from fractal signature analysis (FSA), 2) covariates (age, gender, and body mass index), and 3) texture parameters combined with covariates.

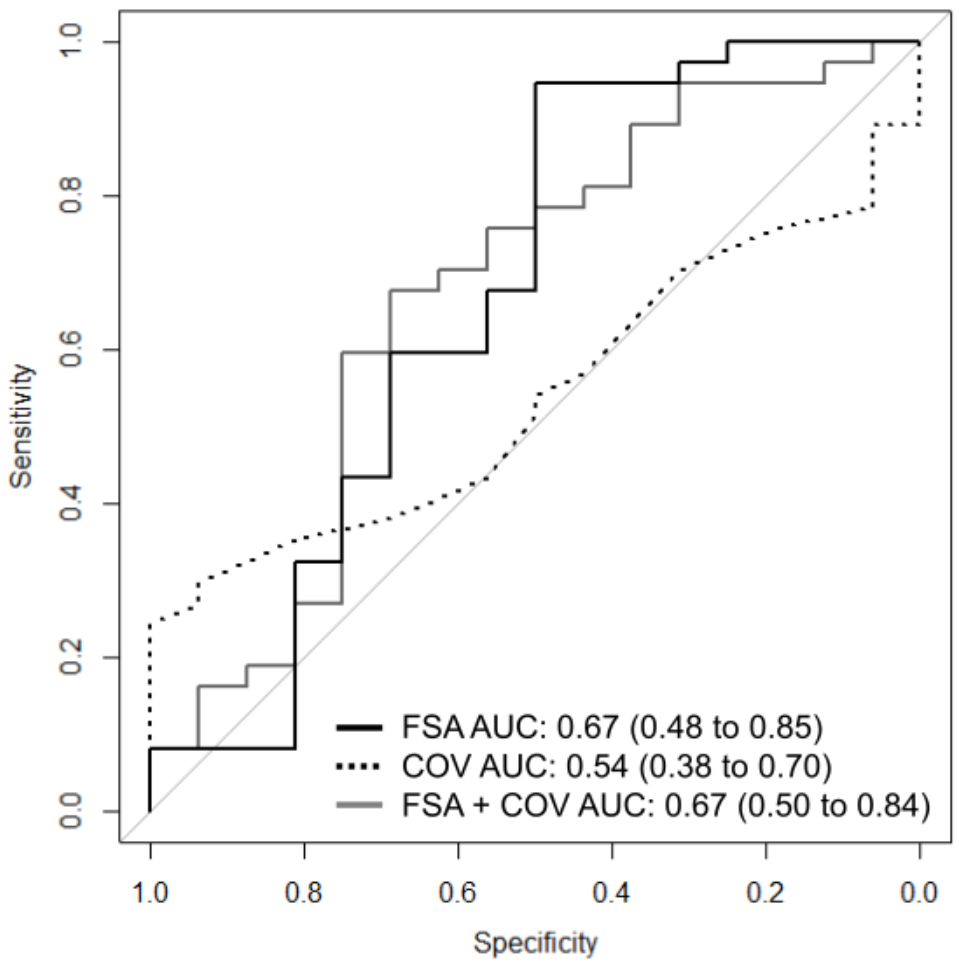

**Figure 6.** Receiver operating characteristics (ROC) curves and their respective area under the curve (AUC) values for predicting incident rHOA (KL ≥ 2) or THR among subjects with KL1 grade at baseline using 1) texture parameters from fractal signature analysis (FSA), 2) covariates (age, gender, body mass index), and 3) texture parameters combined with covariates.

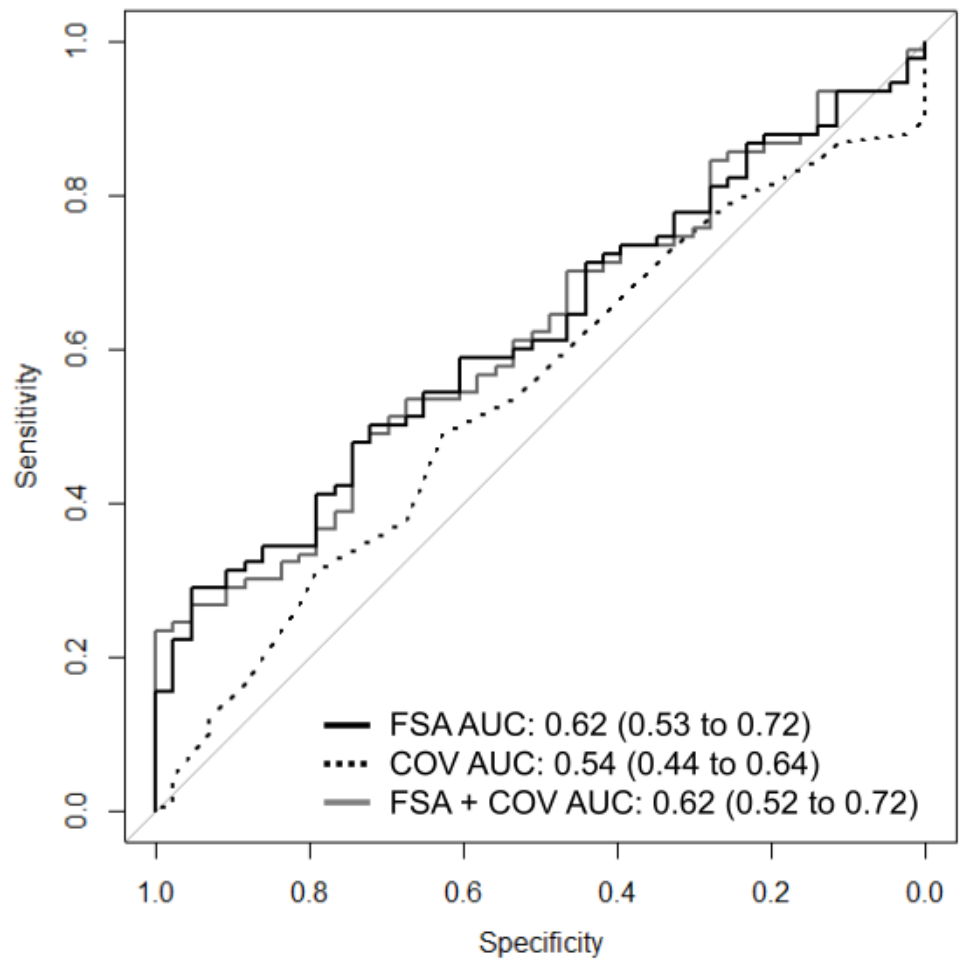

**Figure 7.** Receiver operating characteristics (ROC) curves and their respective area under the curve (AUC) values for predicting incident joint space narrowing (JSN) among subjects with JSN grade 0 at baseline using 1) texture parameters from fractal signature analysis (FSA), 2) covariates (age, gender, body mass index), and 3) texture parameters combined with covariates.

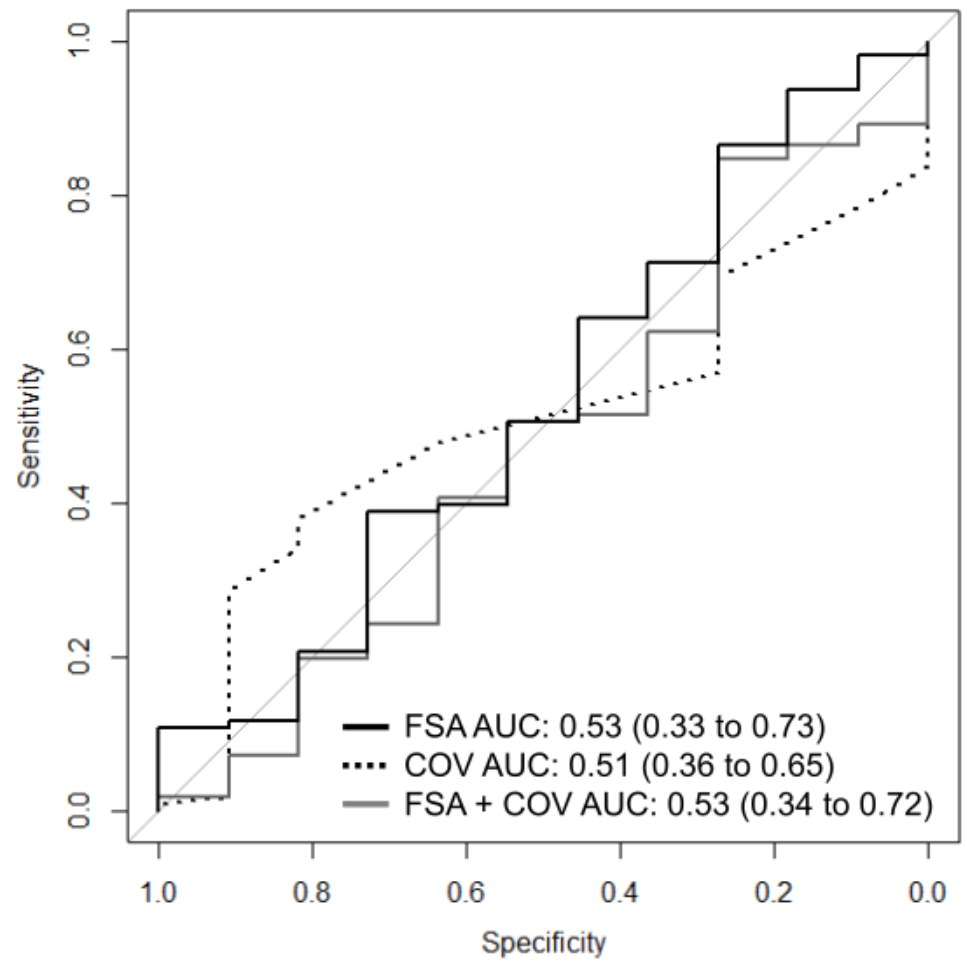

**Figure 8.** Receiver operating characteristics (ROC) curves and their respective area under the curve (AUC) values for predicting incident osteophytes (OST) among subjects with OST grade 0 at baseline using 1) texture parameters from fractal signature analysis (FSA), 2) covariates (age, gender, body mass index), and 3) texture parameters combined with covariates.